\documentclass[reprint,secnumarabic,amssymb,amsmath,aip,cha,groupedaddress,frontmatterverbose]{revtex4-1}
\usepackage{graphicx}

\usepackage{amsmath}
\usepackage{booktabs}
\usepackage{color}

\begin{document}

\title{Efficient thermal energy harvesting  using nanoscale magnetoelectric heterostructures}

\author{S. R. Etesami and J. Berakdar}
\address{Institut f\"ur Physik, Martin-Luther-Universit\"at Halle-Wittenberg, 06099 Halle, Germany}

\begin{abstract}
	Thermomechanical cycles with a ferroelectric working substance convert heat to electrical energy. As shown here, magnetoelectrically coupled ferroelectric/ferromangtic composites (also called multiferroics) add new functionalities and allow for an efficient thermal energy harvesting at  room temperature by exploiting the pyroelectric effect. By virtue of the magnetoelectric coupling, external electric and magnetic fields can steer the operation of these heat engines. Our theoretical predictions are based on a combination of 	 Landau-Khalatnikov-Tani approach (with a Ginzburg-Landau-Devonshire potential) to simulate the ferroelectric dynamics coupled to the magnetic dynamics. The latter is treated via the electric-polarization-dependent Landau-Lifshitz-Gilbert equation. Performing an adapted Olsen cycle we show that a multiferroic working substance is potentially much more superior to sole ferroelectrics, as far as  thermal energy harvesting using pyroelectric effect is concerned. Our proposal holds promise not only for low-energy consuming devices but also for cooling technology.
\end{abstract}
\date{\today}

\maketitle

\begin{center}
	\begin{figure}[t]
		\centering$
		\begin{array}{c}
		\includegraphics[width=5cm]{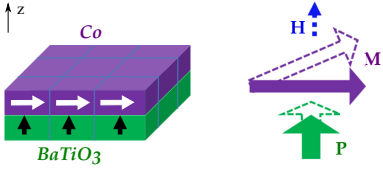}
		\end{array}$
		\caption{\label{schem} Schematics of the magneto-electrically coupled Co/BaTiO$_3$. Both the ferroelectric polarization $\mathbf P$ and the
			magnetization $\mathbf M$ respond to an external magnetic field $\mathbf H$. }
	\end{figure}
\end{center}

It's been known for more than half a century that the temperature-dependency of hysteresis loops in ferroelectrics(FE) can be exploited to convert heat into  electrical energy, known as pyroelectric effect\cite{Childress,Fatuzzo,Ziel,Olsen2,Olsen3,Olsen4,Olsen5,Olsen6,McKinley}. The converse pyroelectric effect is also an established fact called the electrocaloric effect\cite{Olsen1,Mischenko,Vopson,Liu}, which as expected is used in cooling technology. In the quest for environmentally friendly pyroelectric devices that have low energy consumption, multiple functionalities, and being
amenable to integration in nano circuits, we explore in this work the potential of engineered nanoscale multiferroic structures for harvesting waste heat. In particular, we focus on two-phase multiferroic layered structures consisting of a thin layer of the prototypical ferroelectric BaTiO$_3$ (BTO) deposited on Co. At room temperatures a strong magnetoelectric (ME) coupling between BTO and Co  \cite{Jedrecy} was observed. This means that the ferromagnetic Co or BTO  respond to an electric ($E$) or magnetic field ($H$), respectively opening thus  new opportunities for controlling and the possibility for enhancing the device operation. Particularly  important  are room temperature devices in which case BTO is in the tetragonal phase \cite{Li,Wang,Wang1}(see also the supplementary material\cite{supp}, Fig. S2).
%
To exploit the pyroelectric effect to generate electricity, different thermal-electrical cycles were   proposed. We perform the Olsen cycle\cite{Olsen4,Olsen6}: The core idea of the pyroelectric engine is the temperature-dependency of the hysteresis loop (or in other words polarization $P$) since it provides us with the opportunity to create clockwise cycles in $E-P$ space (see the supplementary material\cite{supp}, Fig. S1). The area enclosed by the clockwise cycles determines the amount of harvested thermal energy as
\begin{equation}
	\label{electrical_energy} \varepsilon=V_{FE}\oint PdE.
\end{equation}
Viewing  the ferroelectric (FE) to be consisting  of building blocks (cells or domains)  each with a volume $a^3$  (see Fig.~\ref{schem}) then the FE volume is $V_{FE}=\sum_{\textbf{n}}a^3$. $V_{FE}P=\sum_{\textbf{n}}a^3P_{\textbf{n}}$ is the total charge displacement.
Experimentally, the pyroelectric effect is observed as a flow of an electric charge to and from the surface of FE material. Therefore to increase the efficiency of the pyroelectric engines the dissipated  electrical energy (leakage)  must be decreased as much as possible.
In principle, there are two main sources for dissipation here, hysteresis and Joule heating. Joule heating stems from the finite resistance of FE material, however it can be circumvented by squeezing the cycling time to be much smaller than the characteristic time of the system\cite{Olsen4} $\tau\sim \varepsilon_r\varepsilon_0\varrho\simeq 1$ [s] where $\varepsilon_0\sim10^{-11}$ [AsV$^{-1}$m$^{-1}$], $\varepsilon_r\sim10^4$ and $\varrho\sim10^7$ [VmA$^{-1}$] are the vacuum permittivity, relative permittivity and resistivity of BTO around room temperature, respectively. In our simulations the cycling time is chosen much smaller than $\tau$ so that the relative loss due to charging and discharging is ignorable.

The inevitable energy loss for thermodynamic engines which is manifested in the second law of thermodynamics, determines the maximum possible efficiency achieved by a heat engine, known as Carnot efficiency $\eta^c=\left(1-\frac{T_{cold}}{T_{hot}}\right)$, although approaching this efficiency would be a challenge too.  For example in one of the pioneering experiments on PZST(Pb$_{0.99}$Nb$_{0.02}$(Zr$_{0.68}$,Sn$_{0.25}$,Ti$_{0.07}$)$_{0.98}$O$_3$), for a temperature span of $20$ [K], a pyroelectric efficiency of $\eta^p\approx0.2\%$ was measured while the Carnot efficiency was $\eta^c\approx5\%$\cite{Olsen3,Olsen4}. Theoretically, the efficiency of pyroelectric engines at room temperature is also known to be less than $1\%$. Such a small efficiency stems from the fact that the energy required to increase the temperature of the lattice is nearly always much larger than the energy required to \emph{destroy} part of the polarization, thus releasing electric charges\cite{Fatuzzo}. In our case of study the pyroelectric-engine efficiency is evaluated as
\begin{equation}
	\label{engin_efficieny} \eta^p=\frac{\varepsilon}{Q+w^{ME}},
\end{equation}
where $Q=c\rho V_{FE}\left(T_{hot}-T_{cold}\right)$ is the thermal energy (heat) pumped to the FE material with specific heat $c\sim450$ [Jkg$^{-1}$K$^{-1}$] and mass density of $\rho\sim6000$ [kgm$^{-3}$] at room temperature for BTO\cite{Davitadze}. The temperature span is taken  large enough  so that the pumped transition heat as to have a constant temperature in isothermal processes, is neglegible\cite{Fatuzzo,Olsen4,Olsen6}.
The second term in the denominator is the ME-mediated work done \emph{on} the FE subsystem
\begin{equation}
	\label{work} w^{ME}=\sum_{\textbf{n}}w^{ME}_{\textbf{n}}=-\sum_{\textbf{n}}a^3\oint\lambda P_{\textbf{n}}dM^z_{\textbf{n}},
\end{equation}
where $M^z_{\textbf{n}}$ and $\lambda$ are the magnetization and ME-coupling constant respectively, and $\textbf{n}=(n_x,n_y)$ counts the building blocks of the FE (or ferromagnet(FM)) layer each with volume $a^3$ ($V_{FE}=V_{FM}=\sum_{\textbf{n}}a^3$). This work stems from the ME-mediated effective electric field $E^{ME}_{\textbf{n}}=-\frac{\partial f^{ME}}{\partial P_{\textbf{n}}}=\lambda M^z_{\textbf{n}}$, where $f^{ME}=-\lambda M^z_{\textbf{n}}P_{\textbf{n}}$ is the contribution of ME coupling to total free energy density. The negative $w^{ME}$ means the FE subsystem rejects heat into the FM subsystem. The current study  of using ME coupling to affect the pyroelectric effect is, to our knowledge, the first of its kind. However, in spirit, it is similar
to the experiment using only FE material  in Ref.~\cite{McKinley} and  using compressive stress to stretch  the Olsen cycle and enhance the pyroelectric effect. This aspect is in sofar interesting for future studies, as strain mediated ME coupling is indeed well-established, in addition to other coupling mechanisms \cite{Vaz,Valencia,Jia,Jia2,Janolin,Duan,Duan2,Zhao,Sahoo,Jia3}.

\begin{center}
	\begin{figure}[t]
		\centering$
		\begin{array}{cc}
		\includegraphics[width=3.7cm]{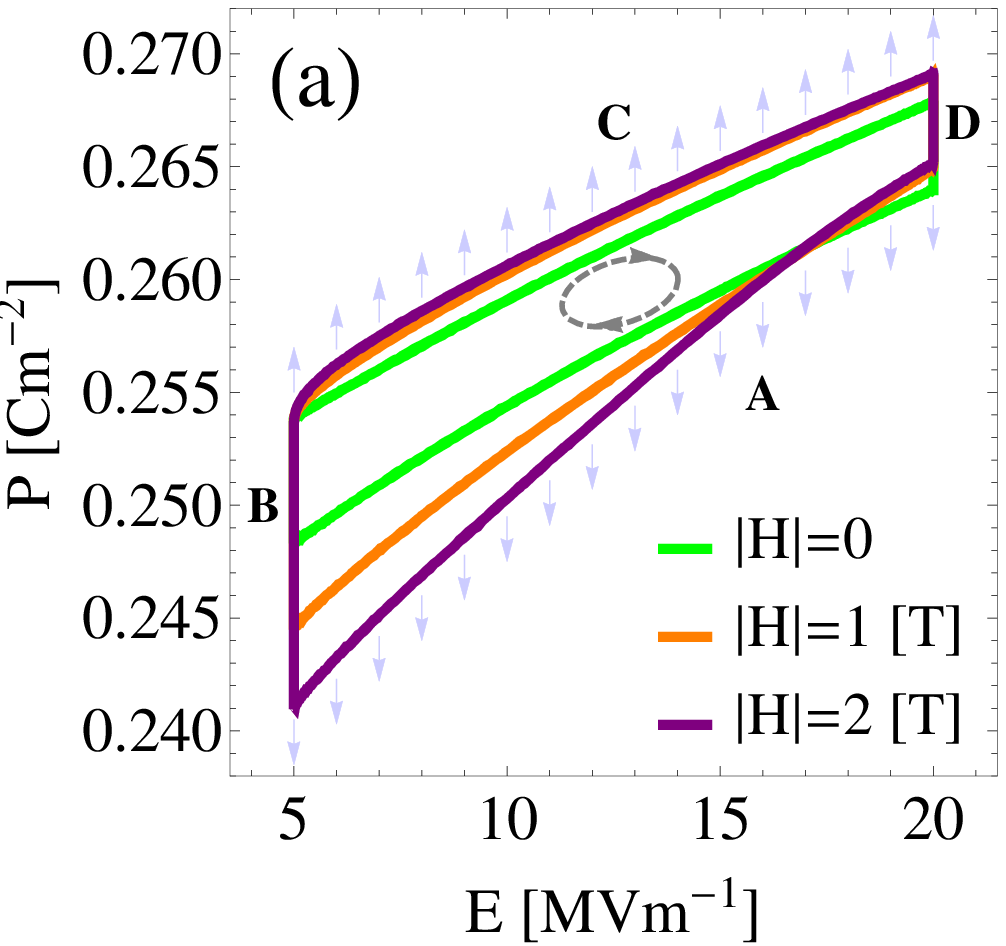}&\includegraphics[width=3.7cm]{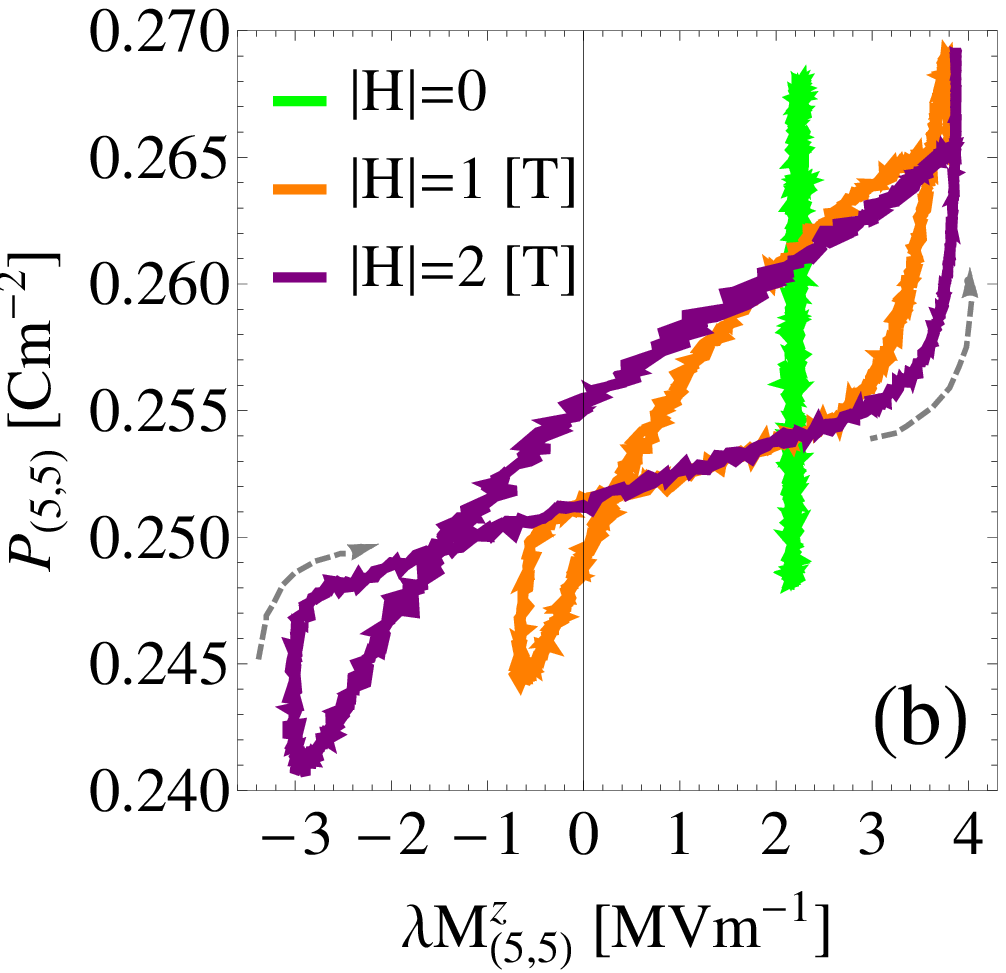}
		\end{array}$
		\caption{\label{olsen_loop} (a) The Olsen cycles for a composite of $10\times10$ FM/FE cells are shown. The cycles consist of four processes \textbf{A}: isothermal discharging at high temperature, \textbf{B}: isoelectric field cooling at low electric field, \textbf{C}: isothermal charging at low temperature, \textbf{D}: isoelectric field heating at high electric field. The area inside the cycles is the produced electrical energy per cubic meter $\varepsilon/V_{FE}=\oint PdE$. The small arrows show the direction of the external magnetic field ($H$) which during \textbf{A}  is directed along -z, and for \textbf{C} it is directed along +z. As can be seen, this oscillatory magnetic field can be used to enhance the area enclosed by the loops. This corresponds to an enhancement of the produced electrical energy. (b) The polarization versus ME-mediated electric field ($E^{ME}_{\textbf{n}}=\lambda M^z_{\textbf{n}}$) at site $\textbf{n}=(5,5)$ is shown. The asymmetry stems from the fact that the magnetization is not solely in x-y plane but it is tilted to +z due to the uniaxial anisotropy and ME coupling (see the supplementary material\cite{supp}, Fig. S3). The area enclosed by the loops determines the ME-mediated work per cubic meter done on the corresponding FE cell. To evaluate the total work done on the whole FE subsystem, the contribution of all other cells must be taken into account $w^{ME}=\sum_{\textbf{n}}w^{ME}_{\textbf{n}}$ (Eq.~(\ref{work})). The calculations are performed at room temperature ($T_{cold}=290$ [K], $T_{hot}=310$ [K]) for $\lambda=2.7$ [sF$^{-1}$] (for more details see the supplementary material\cite{supp}, Fig. S3).}
	\end{figure}
\end{center}

We use the eight-order temperature-dependence Ginzburg-Landau-Devonshire(GLD) potential to simulate BTO. The Landau-Devonshire model has already successfully been used to simulate BTO phase diagram\cite{Li,Wang,Wang1}(see also the supplementary material\cite{supp}, Fig. S2) which in fact makes the pyroelectric simulations feasible and reliable to be realized in experiment. Our model is a FE layer (2D system in x-y plane) including $N\times N$ cells. To address nonhomogeneous multi-domain FE state the Ginzburg gradient($f^G$) is added to the Landau-Devonshire potential\cite{Hlinka,Hlinka1,Marton1}. At room temperature which is our interest, the BTO is in tetragonal phase. In such a case and under an external electric field the GLD potential reads :
\begin{equation}
	\label{tetragonal}
	\begin{split}
		f^{FE}=f^G+\sum_{\textbf{n}}&\alpha_1(T_\textbf{n}) P_{\textbf{n}}^2+\alpha_{11} P_{\textbf{n}}^4+\alpha_{111} P_{\textbf{n}}^6\\
		+&\alpha_{1111} P_{\textbf{n}}^8-E P_{\textbf{n}},\\
		f^G=\sum_{\textbf{n}}&\frac{G_{11}}{2a^2}(P_{(n_x+1,n_y)}-P_{(n_x,n_y)})^2\\
		+&\frac{G_{11}}{2a^2}(P_{(n_x,n_y+1)}-P_{(n_x,n_y)})^2,
	\end{split}
\end{equation}
in which the non-zero component of polarization and external electric field are along z. $G_{11}=51\times10^{-11}$ [C$^{-2}$m$^3$J] determines the strength of the coupling between BTO domains at room temperature\cite{Hlinka}. The Landau-Devonshire potential coefficients\cite{Wang,Wang1,Wang2,Li} are given in the supplementary material\cite{supp}, TABLE I. Temperature is introduced into the system via the potential coefficient $\alpha_1$ and via the noise added to the non-equilibrium effective field $E^{eff}_{\textbf{n}}=-\frac{\partial f^{FE}}{\partial P_{\textbf{n}}}-\gamma_v\frac{dP_{\textbf{n}}}{dt}+\lambda M^z_{\textbf{n}}+\eta_{\textbf{n}}(t)$ with autocorrelation function of $\langle\eta_{\textbf{n}}(t)\eta_{\textbf{m}}(t')\rangle=\frac{2\gamma_vk_BT_{\textbf{n}}}{a^3}\delta_{{\textbf{n}}{\textbf{m}}}\delta(t-t')$, where $\gamma_v\sim2.5\times10^{-5}$ [VmsC$^{-1}$] is the internal resistivity (the inverse of kinetic coefficient)\cite{Sivasubramanian,Marton,Nambu,Chotorlishvili2}. The dynamic of polarization is given by the extended Landau-Khalatnikov-Tani model $\alpha_0\frac{d^2P_{\textbf{n}}}{dt^2}=E^{eff}_{\textbf{n}}$, where $\alpha_0$ is related to the plasma frequency $\omega_0^2=(\frac{G_{11}}{a^2})\alpha_0^{-1}\sim10^{24}$ [s$^{-2}$]\cite{Sivasubramanian,Servoin,Hlinka2,Chotorlishvili,Chotorlishvili2}.

As for FE layer, the FM layer includes $N\times N$ cells in x-y plane. The corresponding free energy density reads :
\begin{equation}
	\label{FM_energy_density}
	\begin{split}
    f^{FM}=\sum_{\textbf{n}}&\frac{k}{M_S^2}\left(M_S^2-(M_{\textbf{n}}^z)^2\right)-\frac{A}{a^2M_S^2}\vec{M}_{\textbf{n}}\cdot\vec{M}_{\textbf{n}'}\\
		+&\frac{1}{2}\mu_0(M_{\textbf{n}}^z)^2-HM_{\textbf{n}}^z,
	\end{split}
\end{equation}
where $H$ is the external magnetic field along z, $k=410$ [kJm$^{-3}$] is the uniaxial anisotropy, $M_S=1.44$ [MAm$^{-1}$] is the saturation magnetization, $A=31$ [pJm$^{-1}$] is the exchange stiffness, $\textbf{n}'$ represents the nearest neighbors and the third term stands for the shape anisotropy with permeability of free space $\mu_0=4\pi\times10^{-7}$ [TmA$^{-1}$]\cite{Coey}. Due to the Cobalt high Curie temperature $T_C=1360$ [K] and the small temperature span around room temperature used in this study (290-310 [K]), the temperature-dependency of coefficients in the FM free energy density is dismissed. Moreover in such range of temperatures the magnitude of the magnetization can reasonably be assumed to be conserved and therefore the dynamic of the magnetization is transversal and can be well described by Landau-Lifshitz-Gilbert(LLG) equation $\frac{\partial}{\partial t}\vec{M}_{\textbf{n}}=-\frac{\gamma}{1+\alpha^2}\vec{M}_{\textbf{n}}\times\left[\vec{H}_{\textbf{n}}^{eff}+\frac{\alpha}{M_S}\vec{M}_{\textbf{n}}\times\vec{H}_{\textbf{n}}^{eff}\right]$, where $\gamma=1.76\times10^{11}$ [T$^{-1}$s$^{-1}$] and $\alpha=0.01$ are the gyromagnetic ratio and Gilbert damping\cite{Barati} respectively. $\vec{H}_{\textbf{n}}^{eff}=-\partial f^{FM}/\partial \vec{M}_{\textbf{n}}+\lambda P_{\textbf{n}}+\vec{\xi}_{\textbf{n}}$ is the effective magnetic field which includes a stochastic field $\vec{\xi}_{\textbf{n}}$ to take into account the thermal fluctuations with the following autocorrelation function $\langle\xi_{i\textbf{n}}(t)\xi_{j\textbf{m}}(t')\rangle=\frac{2\alpha k_B T_{\textbf{n}}}{\gamma M_Sa^3}\delta_{ij}\delta_{{\textbf{n}}{\textbf{m}}}\delta(t-t')$, where the indices $i,j=x,y,z$ denote Cartesian components\cite{Chotorlishvili2,Etesami1,Etesami2}.

To perform Olsen cycle, the system must pass through four steps, as illustrated in Fig.~\ref{olsen_loop} (see also the supplementary material\cite{supp}, Fig. S3). The Olsen cycle includes an isothermal discharging at high temperature(\textbf{A})  followed by a charging process under a constant electric field by cooling the FE material(\textbf{B}). Next, there is an isothermal charging at low temperature(\textbf{C}) and a discharging process, under a constant electric field, by heating FE material(\textbf{D}). At this stage the system is in its initial stage and the Olsen cycle has been accomplished. The area enclosed by the cycle determines the produced electrical energy per cubic meter(Fig.~\ref{olsen_loop}a). To stretch the cycle to increase the produced electrical energy, an oscillatory magnetic field is applied to the composite. During the cold isothermal process (\textbf{C}) the magnetic field is directed in +z and during the hot isothermal process (\textbf{A}) it is directed in -z, as indicated by arrows in Fig.~\ref{olsen_loop}a. This external magnetic field pulls the magnetization towards its own direction, and as a result stretches the polarization due to ME coupling. Of course this enhancement in producing electrical energy comes at
the cost of doing some work on the FE subsystem which must be taken into account to evaluate the engine efficiency (see Eq.~(\ref{engin_efficieny}) and Fig.~\ref{olsen_loop}b).
\begin{center}
	\begin{figure}[t]
		\centering
		\includegraphics[width=\columnwidth]{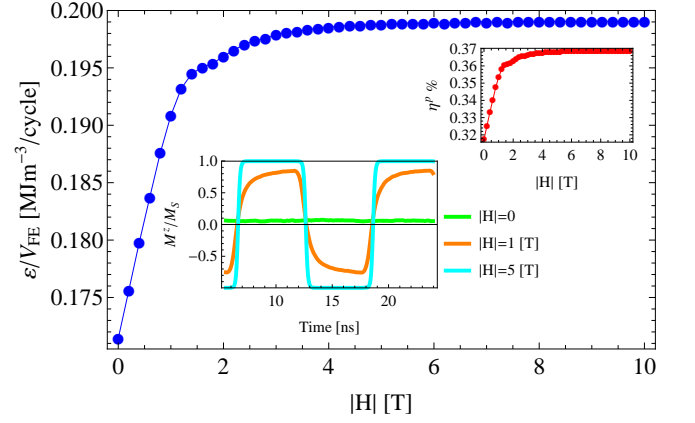}
		\caption{\label{diff_B} Figure shows the produced electrical energy per cubic meter per cycle ($\varepsilon/V_{FE}$) and the corresponding efficiency ($\eta^p$) versus the amplitude of the oscillatory magnetic field ($|H|$). The inset demonstrates that for large amplitudes, the z-component of the magnetization saturates which means a saturation of the ME-mediated electric field ($E^{ME}_{\textbf{n}}=\lambda M^z_{\textbf{n}}$) and so a saturation of the produced electrical energy. The system is at room temperature ($T_{cold}=290$ [K], $T_{hot}=310$ [K]), the electric fields are taken as $E_{low}=5$ [MVm$^{-1}$] and $E_{high}=50$ [MVm$^{-1}$], the cycling time is $12$ [ns] and ME coupling kept at $\lambda=0.27$ [sF$^{-1}$].}
	\end{figure}
\end{center}
\begin{center}
	\begin{figure}[t]
		\centering$
		\begin{array}{c}
		\includegraphics[width=\columnwidth]{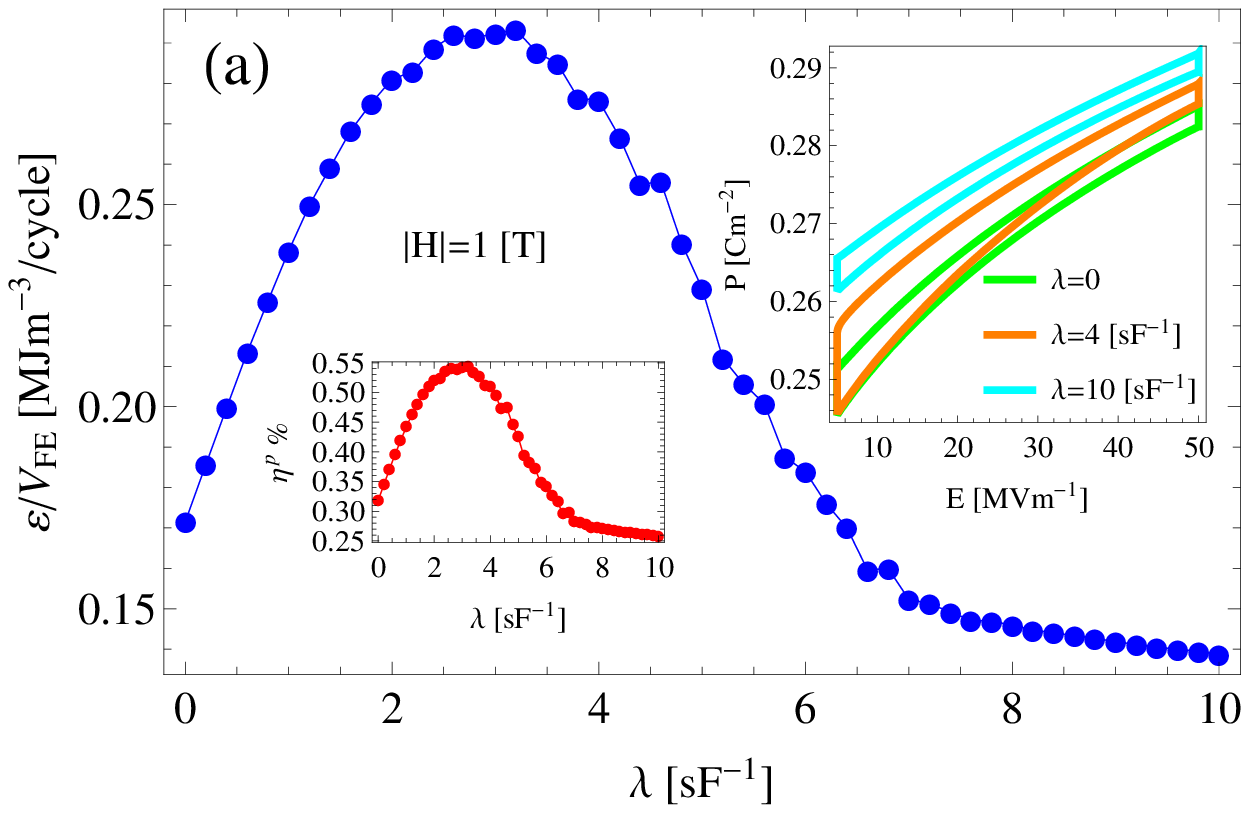}\\
		\includegraphics[width=\columnwidth]{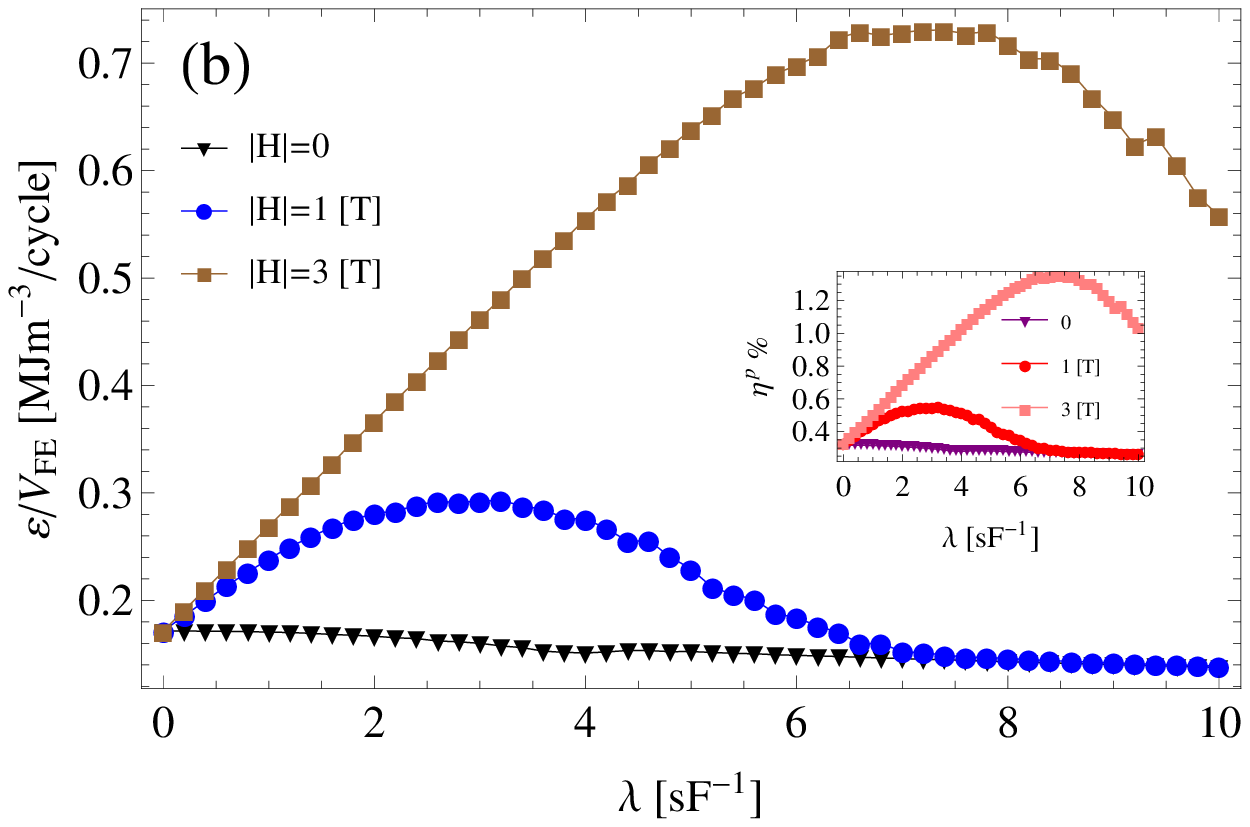}
		\end{array}$
		\caption{\label{diff_g} (a) The produced electrical energy per cubic meter per cycle ($\varepsilon/V_{FE}$) and the corresponding efficiency ($\eta^p$) versus ME coupling ($\lambda$) is shown. The inset shows that increasing ME coupling is not necessarily favorable for the Olsen cycle. (b) The produced electrical energy per cubic meter per cycle ($\varepsilon/V_{FE}$) and the corresponding efficiency ($\eta^p$) versus ME coupling ($\lambda$) for different value of $|H|$ are shown. The system is at room temperature ($T_{cold}=290$ [K], $T_{hot}=310$ [K]), $E_{low}=5$ [MVm$^{-1}$], $E_{high}=50$ [MVm$^{-1}$] and the cycling time is $12$ [ns].}
	\end{figure}
\end{center}

Following the given procedure to perform the adapted Olsen cycle, in Fig.~\ref{diff_B} the effect of the amplitude of the oscillatory magnetic field on produced electrical energy and the efficiency is shown. Since magnetic field solely interact with magnetization, it must be mediated by ME-mediated electric field $E^{ME}_{\textbf{n}}=\lambda M^z_{\textbf{n}}$ to enhance pyroelectric effect. It means that for the oscillatory magnetic fields with high amplitudes in which the z-component of the magnetization is saturated ($\uparrow |H|~\Rightarrow~M^z_{\textbf{n}}\rightarrow M_S~\Rightarrow~E^{ME}_{\textbf{n}}\rightarrow\lambda M_S$) (see the inset in Fig.~\ref{diff_B}) the effect of magnetic field to enhance the produced electrical energy is also saturated.

When it comes to the ME-coupling effect(Fig.~\ref{diff_g}), its role is twofold. On the one hand, increasing the ME-coupling renders the magnetic field more effective for enhancing the pyroelectric effect. On the other hand,  a strong ME coupling  stabilizes the polarization (see the inset in Fig.~\ref{diff_g}a), thus releasing electric charges becomes  more difficult. Therefore, an optimum value for ME coupling is expected in which the produced electrical energy has its maximum value.

\begin{center}
	\begin{figure}[t]
		\centering
		\includegraphics[width=\columnwidth]{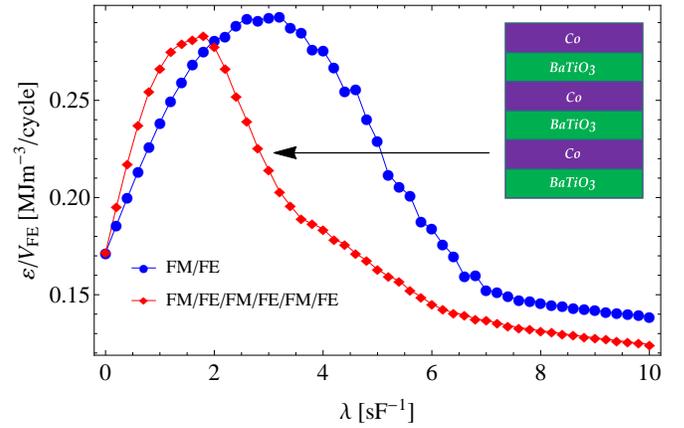}
		\caption{\label{multi_layer} The produced electrical energy per cubic meter per cycle ($\varepsilon/V_{FE}$) versus ME coupling ($\lambda$) for a single FM/FE composite and multi FM/FE composite are compared. The system is at room temperature ($T_{cold}=290$ [K], $T_{hot}=310$ [K]), $|H|=1$ [T], $E_{low}=5$ [MVm$^{-1}$], $E_{high}=50$ [MVm$^{-1}$] and the cycling time is $12$ [ns].}
	\end{figure}
\end{center}

As can be seen in Fig.~\ref{diff_g}b,  increasing the amplitude of the oscillatory magnetic fields softens the polarization easing so the charge release. The optimum ME coupling is expected to move to higher values for higher magnetic fields. Note that, in principle, the ME coupling \textit{alone} is not in favor of pyroelectric effect (Fig.~\ref{diff_g}b: $|H|=0$) since it makes the polarization more resistent to releasing charges. However, implementing an oscillatory magnetic field   enhances the produced electrical energy even to more than $100\%$ as compared  to the case in absence of coupling which makes  multiferroics much more favorable than sole ferroelectrics.
We should mention that  to the best of our knowledge the maximum measured ME coupling is $\lambda=0.27$ [sF$^{-1}$]\cite{Jedrecy} which is much smaller than the range we inspected in our simulations. For  two-phase multiferroics composites  it is possible however to fabricate multiferroics with higher ME coupling due to better techniques and material engineering\cite{Vaz,Vaz2,Vaz3,Wang3,Srinivasan,Nan,Zhai}. Alternatively we can increase the number of layers as FM/FE/FM/FE/FM... to enhance the ME-coupling effect. As can be seen in Fig.~\ref{multi_layer} we can tune the optimum ME coupling by changing the number of layers.

In addition to the strength of ME coupling and the amplitude of external magnetic field, in principle many other parameters can affect the pyroelectric effect such as electric field, cycling time and temperature. However within the scope of this study we were interested in parameters which enhance the produced electrical energy solely within ME coupling. In all  simulations FE and FM layer  consisted of $10\times10$ cells with $a=5$ [nm]; we did not find considerable size effect for larger samples within our model. Moreover, we kept the bath temperatures around room temperature ($T_{cold}=290$ [K], $T_{hot}=310$ [K]) so that the pumped heat per cubic meter per cycle to the FE subsystem and the Carnot efficiency were $Q/V_{FE}\approx54$ [MJm$^{-3}$/cycle] and $\eta^c\approx6.452\%$, respectively.

In summary, due to growing interest and considerable advances in multiferroics on the one hand and the importance of harvesting thermal energy at the nanoscale on the other hand, we proposed a nano-heat engine based on a two-phase multiferroic composite. We exploited the pyroelectric effect and performed an adapted Olsen cycle  showing that it is possible to use such composites to enhance the produced electrical energy at room temperature. We found that a mere ME coupling is not in favor of pyroelectric effect (Fig.~\ref{diff_g}b : $|H|=0$). Implementing an oscillatory magnetic field we can however stretch and expand the enclosed area in the Olsen cycle and reach an optimum value for ME coupling in which the output electrical energy reaches a maximum value (Fig.~\ref{diff_g}). In such a case the presence of ME coupling ($\lambda\neq0$) works substantially in favor of the pyroelectric effect (Fig.~\ref{diff_g}b : $|H|\neq0$) which  might be viewed as an advantage of multiferroics over ferroelectrics. Further, we found the optimum ME coupling is tunable by changing the number of layers (Fig.~\ref{multi_layer}) and the maximum produced electrical energy can be tuned by the amplitude of external magnetic field (Fig.~\ref{diff_g}b). The proposed thermomagnetoelectric cycle adds yet another feature to the technological potentials of multiferroics and holds promise for novel cooling and thermo-sensoric devices.

We thank C.-L. Jia and A. Sukhov for valuable discussions.


\end{document}